\documentclass[paper,12pt]{JHEP3}

\usepackage{amsmath}
\usepackage{epsfig,multicol}
\newcommand{\bc}{\begin{center}}
\newcommand{\ec}{\end{center}}
\newcommand{\be}{\begin{equation}}
\newcommand{\ee}{\end{equation}}
\newcommand{\bea}{\begin{eqnarray}}
\newcommand{\eea}{\end{eqnarray}}
\newcommand{\ba}{\begin{array}}
\newcommand{\ea}{\end{array}}

\newcommand{\bfg}{\begin{figure}[htbp]}
\newcommand{\efg}{\end{figure}}
\def \de {\partial}

\def \m {\mu}
\def \n {\nu}

\def \be {\begin{equation}}
\def \ee {\end{equation}}
\def \bea {\begin{eqnarray}}
\def \eea {\end{eqnarray}}

\def\laq{~\raise 0.4ex\hbox{$<$}\kern -0.8em\lower 0.62
ex\hbox{$\sim$}~}
\def\gaq{~\raise 0.4ex\hbox{$>$}\kern -0.7em\lower 0.62
ex\hbox{$\sim$}~}

\title{Anomalous dimensions and scalar glueball spectroscopy in AdS/QCD}

\author{H. Boschi-Filho, N. R. F. Braga, F. Jugeau, M. A. C. Torres \\
Instituto de F\'{\i}sica,
Universidade Federal do Rio de Janeiro, Caixa Postal 68528, RJ
21941-972 -- Brazil\\
E-mails:  \email{boschi@if.ufrj.br}, \email{braga@if.ufrj.br}, \email{frederic.jugeau@if.ufrj.br},
\email{mtorres@if.ufrj.br}}

\preprint{ }

\abstract{An extended version of the AdS/QCD Soft-Wall model that incorporates
QCD-like anomalous contributions to the dimensions of gauge theory
operators is proposed.
This exploratory approach leads to a relation between scalar glueball masses and
beta functions.
Using this relation, properties of the glueball mass spectroscopy that emerge from phenomenological beta functions proposed in the literature are investigated. The reverse problem is also
considered: starting
from a linear Regge trajectory which fits the lattice glueball masses, beta functions with different asymptotic infrared behaviours are found.
Remarkably, some of them present a fixed point at finite coupling.}


\begin{document}

\section{Introduction}

The idea that string theory and non-abelian gauge theories are related has been proposed 
long ago \cite{'tHooft:1973jz}. More recently, an exact equivalence between string theory
in ten dimensions and gauge theories in four dimensions was discovered 
\cite{Maldacena:1997re,Gubser:1998bc,Witten:1998qj}. This relation holds for string theory
in $AdS_5 \times S^5 $ spacetime and $SU(N)$ Yang Mills theory with large $N$, extended 
${\cal N} = 4$ supersymmetry and conformal invariance.  
This is an example of the AdS/CFT correspondence which also includes other gauge/string 
dualities in different geometries and dimensions. 

In order to apply the idea of gauge/gravity dualities to describe strong interactions, 
it is necessary to break the conformal invariance. 
Presently, an exact dual description of QCD is not known.
However, in the recent years, some important QCD properties have been reproduced from 
phenomenological models based on the AdS/CFT correspondence. Essentially, these so-called AdS/QCD 
models consist in modifying the AdS geometry with the purpose of breaking the conformal invariance. 

An infrared scale in the gauge theory was associated with a localization in the AdS space
in Ref.\cite{Polchinski:2001tt}. 
This way a physical process with an infrared scale in four dimensions 
is mapped into a region of the AdS space. Using this idea, the correct high energy 
scaling of hadronic  amplitudes for fixed angle scattering was found. 
This experimentally observed scaling was reproduced by QCD long before in 
\cite{Matveev:1973ra}. 

Putting forward the idea of Ref.\cite{Polchinski:2001tt} to relate
AdS/CFT and QCD, in Ref. \cite{BoschiFilho:2002ta} the scalar
glueball spectroscopy was studied using an AdS slice.  The size of the slice is related to  $\Lambda_{QCD}$. Considering boundary conditions in the AdS slice, normalizable modes
for the scalar bulk fields dual to scalar glueballs with a discrete spectrum were found.
The approach of using an AdS slice to investigate hadronic properties was then called AdS/QCD
Hard-Wall model and applied to other particles (see \emph{e.g.} \cite{deTeramond:2005su}). 

It was then realized that the Regge trajectories that can be obtained from the Hard-Wall model
are not linear. Then, another AdS/QCD model was proposed to give linear trajectories, especially for vector mesons. This was done with the introduction of a non-dynamical dilaton background field that 
plays the role of a smooth cut-off in the AdS spacetime \cite{Karch:2006pv}. 
This is called Soft-Wall model and was also applied to study scalar particle properties \cite{Colangelo:2007pt,Forkel}. 

The AdS/CFT correspondence deals (in its weakest version) with a strong Yang-Mills coupling. However, a complete holographic description of QCD should incorporate its infrared slavery as well as asymptotic freedom properties. Regarding the importance of the issue, it is worthwhile to explore dual mechanisms able to implement a non-vanishing beta function in the AdS/QCD framework. In this respect, an interesting string-like approach to holographic QCD appeared in \cite{Gursoy:2007cb,Gursoy:2007er} (see also \cite{Gursoy:2010fj} for a review). In these models, (dynamical) dilaton potentials are related to QCD-like beta functions and glueball mass spectra were obtained in agreement with lattice results.

In the AdS/CFT correspondence, the masses of supergravity fields are related to the scaling 
dimensions of the local gauge-invariant dual operators. This relation is derived from the ultraviolet asymptotic behaviour of the bulk-to-boundary propagators \cite{Gubser:1998bc,Witten:1998qj}. Since the gauge theory is conformal, the beta function vanishes and the scaling dimensions do not get anomalous contribution keeping their 
canonical - or classical - dimensions. In AdS/QCD, it is usually assumed that the relation between the masses of 
the bulk fields and the dimensions of the boundary operators is the same 
as given by AdS/CFT (for an exception, see \emph{e.g.} \cite{vega}).

The exploratory approach we follow in this paper aims especially at improving the phenomenological AdS/QCD Soft-Wall model for the case of scalar glueballs. The motivation is that this model predicts masses systematically smaller than lattice results \cite{Colangelo:2007pt}. We will explore the effects on the masses due to the full dimension of the glueball operator. The anomalous contribution implies a modification of the mass of the dual $5d$ supergravity field which, in turn, gives rise to a modification of the $4d$ mass spectrum of the scalar glueballs. We will consider some possible QCD-like beta functions with different infrared behaviours and investigate the main features of the corresponding glueball mass spectra. 

The paper is organized as follows: in section 2, we review lattice results, discuss the Soft-Wall model and calculate the anomalous dimension of the scalar glueball operator. In section 3, we study the scalar glueball mass spectrum from three different non-perturbative QCD beta function models  considered in the literature. Section 4 is devoted to constructing beta functions able to reproduce linear Regge trajectories of lattice glueball masses.

%%%%%%%%%%%%%%%%%%%%%%%%%%%%%%%%%%%%%%%%%%%%%%%%%%%%%%%%%%%%%%%%%%%%%%%%%%%%%%%%%%%%%%%%%%%%%

\section{Scalar Glueballs} 

\subsection{Lattice results}

The glueball mass spectrum has been the subject of numerous studies in lattice QCD \cite{Meyer:2004gx,Morningstar:1999rf,Chen:2005mg,Lucini:2001ej}, which have 
provided estimates for the ground-state and a few excited states as shown in Table \ref{tab:latticeresults} (for general reviews, see for instance \cite{Mathieu:2008me,Klempt:2007cp}).
For comparison, we also show the ratios of the glueball masses:

\begin{equation}
\label{ratios}
R_1\,=\,\frac {M_{0^{++\ast}}}{ M_{0^{++}}}\,; \quad R_2\,=\,\frac {M_{0^{++\ast\ast}}}{ M_{0^{++}}}\,; \quad R_3\,=\,\frac {M_{0^{++\ast\ast\ast}}}{ M_{0^{++}}}\,,
\end{equation}
\noindent   and the corresponding systematic errors since the errors coming from the string tension do not contribute to the ratios.

\begin{table}[!h]
%\centering
\begin{center}
  \begin{tabular}{| c || c || c | c || c | c |}
    \hline\hline
     & Ref.  \cite{Meyer:2004gx} & Ref.  \cite{Morningstar:1999rf} & Ref.   
    \cite{Chen:2005mg} & \multicolumn{2}{| c |}{ Ref. \cite{Lucini:2001ej}} \\ \hline
            $J^{PC}$        &     $N_c=3$       & \multicolumn{2}{|c||}{ $N_c=3$,  \textrm{anisotropic lattice}} 
     & $N_c=3$ & $N_c \to \infty$ \\ \hline
    $0^{++}$ & {1.475}(30)(65) & 1.730(50)(80) & 1.710(50)(80) & 1.58(11) & 1.48(07) \\ \hline
    $0^{++\ast}$ & {2.755}(70)(120) & 2.670(180)(130) & & 2.75(35) & 2.83(22) \\ \hline
    $0^{++\ast\ast}$ & {3.370}(100)(150) & & & & \\ \hline
    $0^{++\ast\ast\ast}$ & {3.990}(210)(180) & & & & \\
\hline\hline 
$R_1$ &  1.87(8) & 1.54(15)& & 1.74(34) & 1.91(24)
\\ \hline
$R_2$ & 2.28(11)  & & & & 
\\ \hline 
$R_3$ & 2.71(20)  & & & & 
\\ \hline\hline 
  \end{tabular}
\end{center}
\caption{Lattice scalar glueball mass spectra in GeV    and mass ratios. The errors are shown in parenthesis and described in the text.}
\label{tab:latticeresults}
\end{table}

In the second column, we show the results of Ref.\cite{Meyer:2004gx} computed for pure $SU(3)_c$
lattice gauge theory. The first number in parenthesis is the statistical error stemming from the continuum-limit 
extrapolation while the second error accounts for the uncertainty in the string tension $\sigma$. 

In the third and fourth columns of Table \ref{tab:latticeresults}, we show results from 
Refs.\cite{Morningstar:1999rf,Chen:2005mg} where anisotropic lattices have 
been used with different temporal and spatial spacings. 
The first error comes from the combined uncertainties from the continuum-limit extrapolation
and from the anisotropy, the second from the uncertainty in a  
hadronic length scale playing a role similar to the string tension.

It is worth pointing out that most of the investigations in lattice consider 
$SU(N_c)$ theories at finite $N_c$. However, in the AdS/CFT correspondence, the Yang-Mills theory is in 
the large $N_c$ limit. In order to compare the large-$N_c$ gauge theory to its finite counterpart, Ref.\cite{Lucini:2001ej} calculated the lightest and the first excited scalar glueball masses in gauge theories 
for increasing $ N_c $ and it was found a mass difference for the glueballs of about only 
$5\%$ between the $N_c=3$ and the large-$N_c$ gauge theories (see also \cite{Lucini:2010nv}).  These results are shown 
in the last two columns of Table \ref{tab:latticeresults} where we used the mean value $\sqrt{\sigma}=440$ MeV of \cite{Meyer:2004gx}. 

Note also that unquenched lattice QCD only provides the ground-state scalar glueball mass, usually suffering from severe computational difficulties (lack of high statistics, coarse lattice spacing, etc) \cite{Bali:2000vr}. Finally, let's mention that unsubtracted QCD spectral sum rules in pure Yang-Mills predicts a ground-state mass around 1.5 GeV \cite{narison}.

%%%%%%%%%%%%%%%%%%%%%%%%%%%%%%%%%%%%%%%%%%%%%%%%%%%%%%%%%%%%%%%%%%%%%%%%%

\subsection{The anomalous dimension of the scalar glueball operator in QCD}

Following \cite{Gubser:2008yx}, the full dimension of the scalar glueball operator can be obtained from the trace anomaly of the QCD energy-momentum tensor \cite{Narison:1989}:
\begin{equation}\label{traceanomaly}
T^\mu_\mu=\frac{\beta(\alpha)}{16\pi\alpha^2}Tr\,G^2
+(1+\gamma_m(\alpha))\sum_{n_f}m_{q_f}\bar{q}_fq_f
\end{equation} 
where the beta function is defined as usual as 
\begin{equation}
\beta(\alpha(\mu)) \equiv \frac{d\alpha(\mu)}{d\ln(\mu)}
\end{equation}
with $\mu$ the renormalization scale, $\alpha\equiv{g_{_{YM}}^{\,2}}/{4\pi}$ and $g_{YM}$ the Yang-Mills coupling constant.

On the other hand, for any operator $\mathcal{O}$, we have the following scaling behaviour:
\begin{equation}\label{scalingbehaviour}
\Delta_{_{\mathcal{O}}} \, \mathcal{O} = -\frac{d\mathcal{O}}{d\ln \mu}
\end{equation}
where the full dimension $\Delta_{_{\mathcal{O}}}=\Delta_{class.}+\gamma(\mu)$ is given in
terms of the classical dimension $\Delta_{class.}$ and the anomalous dimension $\gamma(\mu)$. 
In the following, the fermionic contribution on the \emph{r.h.s} of (\ref{traceanomaly})
will not be considered, since we are interested only in the equation for the operator $Tr\,G^2$. Thus, by taking into account the scalar glueball operator contribution of the QCD trace anomaly, 
Eq.(\ref{scalingbehaviour}) gives: 
\begin{eqnarray}
\Delta_{T_{\m}^{\m}} \left( \frac{\beta(\alpha)}{8\pi\alpha^2} Tr\, G^2 \right) 
&=& - \frac{d}{d\ln \mu} \left( \frac{\beta(\alpha)}{8\pi\alpha^2} Tr\, G^2 \right) \cr \cr
&=& - \big(\beta'(\alpha)-\frac{2}{\alpha}\beta(\alpha)-\Delta_{G^2} \big)\frac{\beta(\alpha)}{8\pi\alpha^2}Tr\,G^2 
\end{eqnarray}
where the prime denotes the derivative with respect to $\alpha$. 
The trace $T^{\m}_{\m}$ scales classically, that means $\Delta_{T_{\m}^{\m}}=4$. 
This finally implies that the scalar glueball operator $Tr\,G^2$ has the full dimension: 
\begin{equation}\label{anondim}
\Delta_{G^2}=4+\beta'(\alpha)-\frac{2}{\alpha}\beta(\alpha)
\end{equation}
which, in terms of the  't Hooft coupling $\lambda\equiv N_c\,g_{YM}^2=4\pi N_c\,\alpha$, reads 
\begin{equation}\label{anondim2}
\Delta_{G^2}=4+\beta'(\lambda)-\frac{2}{\lambda}\beta(\lambda)
\end{equation}
where a prime now denotes a derivative with respect to $\lambda$ and 
\begin{equation}
\beta(\lambda(\mu))=\frac{d\lambda(\mu)}{d\ln(\mu)}\;.
\end{equation}

%%%%%%%%%%%%%%%%%%%%%%%%%%%%%%%%%%%%%%%%%%%%%%%%%%%%%%%%%%%%%%%%%%%%%%%%

\subsection{The Soft-Wall model}\label{massissue}

In the Soft-Wall model \cite{Karch:2006pv}, the dynamics of a massive scalar bulk field $X=X(x,z)$ is governed by the following action \cite{Colangelo:2007pt}:
\begin{equation}
S \,=\,-\frac{1}{k}\int d^{5}x \, \sqrt{-g}\,e^{-\Phi(z)}\, [g^{MN}\partial_M X\partial_N X+m_{AdS}^2X^2]
\label{action}
\end{equation}
\noindent 
where $g$ is the determinant of the metric tensor of $AdS_5 $ given by 
\begin{equation}
ds^2 \equiv g_{MN}dx^M dx^N \, = \, 
\frac{R^2}{z^2}\big(\eta_{\m\n}dx^{\m}dx^{\n}+dz^2\big)\,,
\label{ads}
\end{equation}
with $\eta_{\mu\nu}=\textrm{diag}(-1,+1,+1,+1)$ the Minkowski $4d$ metric. $R$ is the $AdS_5$ radius and $z$ is the holographic coordinate allowed to run from zero to infinity. In the following, we will set $R=k=1$ as they do not enter expressions for the $4d$ particle masses. 

The conformal symmetry is smoothly broken by an infrared cut-off represented by a non-dynamical scalar field (the so-called ``background dilaton field'') chosen as $\Phi(z)=c z^2$ where the parameter $c$ has the dimension of a squared mass. At odds with the Hard-Wall model, this conformal symmetry breaking mechanism allows for linear Regge trajectories. The background field $\Phi(z)$ of the phenomenological Soft-Wall model is not dual to any mode living on the four dimensional boundary spacetime and does not follow from the solution of Einstein equations \cite{csaki1,csaki2}. 

The equation of motion for the bulk field $X(x,z)$ is:
\begin{equation}
\de_z\big(\frac{1}{z^3}e^{-\Phi(z)}\de_z
X\big)+\frac{1}{z^3}e^{-\Phi(z)}\eta^{\m\n}\de_{\m}\de_{\n}X-\frac{1}{z^5}e^{-\Phi(z)}m_{AdS}^2X=0\,.
\end{equation}
\noindent 
Representing the scalar field through a $4d$ Fourier transform $\tilde{X}(q,z)$, one finally gets, under the change of
function $\tilde{X}=e^{B/2}\tilde{Y}$ with $B(z)=\Phi(z)+3\ln(z)$, a $1d$ Schr\"odinger-like equation:
%$$X(x,z)=\int\frac{d^4q}{(2\pi)^4}e^{iq\cdot x}\tilde{X}(q,z)\,,$$ 
%one has:
%\begin{equation}
%\de_z\big(e^{-B(z)}\de_z\tilde{X}\big)-q^2e^{-B(z)}\tilde{X}-\frac{m_{AdS}^2}{z^2}e^{-B(z)}\tilde{X}=0
%\end{equation}
%with $B(z)=\Phi(z)+3\ln(z)$. 
\begin{equation}\label{Schroe}
-\de_z^2\tilde{Y}+V(z)\tilde{Y}=-q^2\tilde{Y}
\end{equation}
with the $5d$ effective potential:
\begin{equation}\label{potencial0}
V(z)=\frac{B'^2}{4}-\frac{B''}{2}+\frac{m_{AdS}^2}{z^2}=
c^2z^2+\frac{15}{4z^2}
+2c +\frac{m_{AdS}^2}{z^2}\,.
\end{equation}
The normalizable solutions of Eq.(\ref{Schroe}) correspond to a discrete spectrum of $4d$ masses 
$q_n^2\equiv - m_n^2$. 

In the Soft-Wall model, one assumes that the $5d$ mass $m_{AdS}$ is related to the scaling dimension of the dual $4d$ boundary operator as given by the AdS/CFT correspondence. For the scalar case, on has:
\begin{equation} \label{AdSmass}
m_{AdS}^2=\Delta(\Delta-4)\,.
\end{equation}

Within this holographic set-up, the scalar glueball has been investigated. It is associated with the local gauge-invariant QCD operator $Tr\,G^2$ defined on the boundary spacetime and which has 
classical dimension $\Delta_{class.}=4$. Using (\ref{AdSmass}), the bulk mass vanishes:
\begin{equation}
m_{AdS}=0\label{AdSmassScalar}
\end{equation}
and the corresponding scalar glueball mass spectrum was found in \cite{Colangelo:2007pt}:
\begin{equation}\label{masssoftwall}
m_{G_n}^2=4c(n+2)\,\,.
\end{equation} 
If one fixes $c$ by the mass spectrum of the vector $\rho$ mesons in AdS/QCD \cite{Karch:2006pv} as 
\begin{equation}
c=0.2325\;\textrm{GeV}^2=(0.482\;\textrm{GeV})^2\,\,,\label{SWc}
\end{equation}
one obtains the results shown in Table \ref{colan}. 

%%%%%%%%%%%%%%%%%%%%%%%%%%%%%%%%%%%%%%%%%%%%

\begin{table}[!h]
\centering
\begin{center}
  \begin{tabular}{| c | c | c | c |}
    \hline\hline
      $0^{++}$ & $0^{++\ast}$ & $0^{++\ast\ast}$ & $0^{++\ast\ast\ast}$  \\ \hline\hline
      1.364 & 1.670 & 1.929 & 2.156 \\ \hline\hline 
  \end{tabular}
\end{center}
\caption{Scalar glueball masses in GeV from the Soft-Wall model \cite{Colangelo:2007pt}.}
\label{colan}
\end{table}

%%%%%%%%%%%%%%%%%%%%%%%%%%%%%%%%%%%%%%%

One can see that these results from the Soft-Wall model are systematically smaller\footnote{With a negative dilaton parameter and keeping the value (\ref{SWc}), one gets even smaller masses $m_{G_n}^2=4|c|(n+1)$ \cite{FJ}.} than the lattice values of glueball masses reviewed in subsection {\bf 2.1}. Furthermore, it is interesting to note that even changing the value of $c$, one can not fit the lattice glueball masses using the relations (\ref{AdSmassScalar}) and (\ref{masssoftwall}). 
The mass ratios defined in Eq. (\ref{ratios}) for the soft wall model are
\begin{equation}
\label{ratiosSoftwall}
R_1\,=\,1.22 ; \quad R_2\,=\,1.41 ; \quad R_3\,=\, 1.58\,.
\end{equation}
\noindent   Note that these values are not in agreement with the lattice results shown in Table
\ref{tab:latticeresults}. 

In the following, we will see how the mass ratios can be improved by the introduction of the anomalous contribution to the dimension of the scalar glueball operator.

%%%%%%%%%%%%%%%%%%%%%%%%%%%%%%%%%%%%%%%%%%%%%%%%%%%%%%%%%%%%%%%%%%%%%%%%

\section{Improved Soft-Wall model and glueball masses}

\subsection{Soft-Wall model with QCD-like anomalous dimensions}

Our investigation relies on the Soft-Wall model reviewed in subsection {\bf 2.3} from which assumptions and results are well understood. In this paper, considering the case of the scalar glueball operator, we will study the main effects on the spectroscopy due to the anomalous contribution of the operator dimension. The anomalous dimension stems from quantum effects and depends on a renormalization scale $\mu$. Since standard AdS/CFT holography implies that the fifth $z$ coordinate is inversely proportional to the $4d$ energy scale $\mu$ \cite{Susskind:1998dq}, the full dimension (\ref{anondim2}) and thus the bulk mass (\ref{AdSmass}) also depend on the holographic $z$ coordinate. 
 
At small $z$, any asymptotically AdS spacetime bulk gravity theory satisfies the AdS/CFT relation (\ref{AdSmass})
exactly. In the Soft-Wall model, one  extrapolates this relation to the large $z$ region, assuming its validity at any energy scale. As a result we write the Eq. (\ref{AdSmass}) with the dimension $\Delta$ as a function of $z$:
\begin{equation}
m_{AdS}^2(z)=\Delta(z)(\Delta(z)-4)
\end{equation}
where the full dimension of the scalar glueball operator $\Delta(z)$ is given by the QCD result (\ref{anondim2}). Note that this procedure has also been followed in the context of finite temperature QCD \cite{Gubser:2008yx}: considering the conformal AdS/CFT relation (\ref{AdSmass}) extended to the case of non-zero anomalous dimensions, the authors approximately reproduced the 2+1 flavour lattice QCD estimate of the squared speed of sound as function of the temperature.
 
Finally, the $z$-dependent bulk mass generalizes the $5d$ Soft-Wall potential (\ref{potencial0}) to  
\begin{eqnarray}\label{potencialanomalous}
V(z)
&=& c^2z^2+\frac{15}{4z^2}+2c+\frac{1}{z^2}\left[4+\beta'(\lambda)-\frac{2}{\lambda}\beta(\lambda)\right]
\left[\beta'(\lambda)-\frac{2}{\lambda}\beta(\lambda)\right]\,\,.
\end{eqnarray}
We will solve  the $1d$-Schr\"odinger-like equation (\ref{Schroe}) with this effective potential for different phenomenological beta functions and examine the properties of the corresponding glueball mass spectra.

%%%%%%%%%%%%%%%%%%%%%%%%%%%%%%%%%%%%%%%%%%%%%%%%%%%%%%%%%%%%%%%%%%%%%%%%

\subsection{Glueball spectroscopy from improved Soft-Wall model}
 
Perturbatively, the QCD beta function is expressed as a power series of the coupling.
Each term of the series comes from a certain loop order. The first and second terms are independent of the renormalization scheme, while the higher order terms are scheme dependent.  
At strong coupling one can not rely on the perturbative expansion of the beta function.

In this section, we consider some effective non perturbative beta functions that could mimic the
infrared behavior of QCD. We will use beta functions discussed in the literature \cite{Zeng:2008sx,Alanen:2009na,Ryttov:2007cx} and adjust their parameters in order to obtain a mass spectrum compatible with the glueball lattice results. 
We also demand that the beta functions
have an ultraviolet perturbative behaviour similar to QCD for small $\lambda$ in 1-loop approximation:
\begin{equation}\label{asymptfreed}
\beta (\lambda) \sim -b_0 \lambda^2
\end{equation}
where $b_0$ is the universal coefficient of the perturbative QCD beta function at leading order:
\begin{equation}
b_0=\frac{1}{8\pi^2}\left(\frac{11}{3}-\frac{2}{9}n_f\right)
\end{equation}
where we will take $n_f=0$ such that $b_0=11/24\pi^2$. Our analysis deals indeed with the lattice study \cite{Meyer:2004gx,Morningstar:1999rf,Chen:2005mg,Lucini:2001ej}
which was able to predict the four lowest-lying scalar glueball masses in pure $SU(3)_c$. 

As a first example of beta function, note that the \emph{r.h.s.} of (\ref{asymptfreed}) was considered in \cite{Zeng:2008sx} when investigating the heavy quark-antiquark interaction potential in some 
Renormalization Group revised AdS/QCD models. 
However, in our framework, such a beta function cannot give additional 
anomalous contribution to the effective potential, since the last term in (\ref{potencialanomalous}) 
vanishes identically in this case. 

In the Soft-Wall model of QCD, 
the fifth coordinate $z$ of the AdS spacetime is identified with $\mu^{-1}$ where $\mu$ is the renormalization group scale (see, \emph {e.g.} \cite{Karch:2006pv,Colangelo:2007pt,Forkel,csaki1} and references therein). Hence, the equation for the beta function becomes in terms of $z$:
\begin{equation}\label{betaf}
\mu\frac{d\lambda(\mu)}{d\mu}=\beta(\lambda(\mu))\;\;\Rightarrow\;\;z\frac{d\lambda(z)}{dz}=-\beta(\lambda(z))
\end{equation}
where the integration constant will be fixed such that $\lambda(z_0)\equiv\lambda_0$ at a particular energy scale $z_0$. 

%%%%%%%%%%%%%%%%%%%%%%%%%%%%%%%%%%%%%%%%%%%%%%%%%%%%%%%%%%%%%%%%

\subsubsection{Beta function with an IR fixed point at finite coupling}

First, let us consider the following beta function \cite{Alanen:2009na}:
\begin{equation}\label{betalambdastar}
\beta(\lambda)=-b_0\lambda^2\left(1-\frac{\lambda}{\lambda_{\ast}}\right) \qquad \qquad  
(\lambda_{\ast}>0)\;.
\end{equation}
This beta function vanishes at the infrared fixed point $\lambda=\lambda_{\ast}$, reproduces the perturbative $\beta(\lambda)\sim-b_0\lambda^2$ at 1-loop order in the ultraviolet
and behaves as $\beta(\lambda)\sim +\lambda^3$ at large coupling. 

The Renormalization Group equation (\ref{betaf}) for this beta function can be exactly solved, finding:
\begin{equation}
\lambda(z)=\frac{\lambda_{\ast}}{1+W\left(\left(\frac{z_0}{z}\right)^{b_0\lambda_{\ast}}\left(\frac{\lambda_{\ast}-\lambda_0}{\lambda_0}\right)e^{\frac{\lambda_{\ast}-\lambda_0}{\lambda_0}}\right)}
\end{equation}
which leads to the proper QCD asymptotic behaviour at short distances when $z$ is close to the boundary: 
\begin{equation}
\lambda(z)\sim-1/(b_0\ln z)\,\,.\label{QCDlike}
\end{equation}
$W(x)$ is the Lambert function and $\lambda(z_0)=\lambda_0$ fixes the integration constant.
%\begin{equation}
%z(\lambda)=z_0 e^{\frac{1}{b_0}\left(\frac{1}{\lambda_0}-\frac{1}{\lambda}\right)}
%\left(\frac{\lambda_{\ast}-\lambda_0}{\lambda_{\ast}-\lambda}\frac{\lambda}{\lambda_0}
%\right)^{\frac{1}{b_0\lambda_{\ast}}}
%\end{equation}
Then, the $5d$ effective potential (\ref{potencialanomalous}) takes the form:
\begin{equation}
V(z)=c^2z^2+\frac{15}{4z^2}+2c+\frac{b_0\lambda_{\ast}}{z^2}\frac{\Big[4\left(1+W\left((\frac{z_0}{z})^{b_0\lambda_{\ast}}(\frac{\lambda_{\ast}-\lambda_0}{\lambda_0})e^{(\frac{\lambda_{\ast}-\lambda_0}{\lambda_0})}\right)\right)^2+b_0\lambda_{\ast}\Big]}{\Big[1+W\left((\frac{z_0}{z})^{b_0\lambda_{\ast}}(\frac{\lambda_{\ast}-\lambda_0}{\lambda_0})e^{(\frac{\lambda_{\ast}-\lambda_0}{\lambda_0})}\right)\Big]^4}
\end{equation}
which behaves in the \emph{infrared} as
\begin{equation}
V(z) \sim c^2z^2+\frac{[15+4b_0\lambda_{\ast}(4+b_0\lambda_{\ast})]}{4z^2}+2c\,\,.
\end{equation}
In other words, in the large $z$ limit (when $\lambda$ goes to $\lambda_{\ast}$), the contribution to the potential stemming from the 
anomalous dimension is subleading with respect to the oscillator-like term $c^2z^2$ 
coming from the background dilaton field. As a result, the Regge-like behaviour of the Soft-Wall mass spectrum $m_n^2\,{\sim}\, n$ is preserved for large enough $n$. Similarly, in the ultraviolet, the anomalous contribution to the potential gives a subleading term with respect to the usual $AdS_5$ term in $1/z^2$:
\begin{equation}
V(z)\sim c^2z^2+\frac{15}{4z^2}+2c+\frac{4}{b_0\lambda_{\ast}}\frac{1}{z^2\ln(\frac{z}{z_0})}\,\,,
\end{equation}  
consistent with the asymptotically conformal behaviour of the gauge theory. Thus, the anomalous contribution in (\ref{potencialanomalous}) consists, for the beta function (\ref{betalambdastar}), in modifying the $5d$ potential for intermediate values of $z$ only and provides, as it will be shown below, a way of solving the issue discussed at the end of subsection {\bf 2.3}.

In order to do the numerical analysis, we employ a standard shooting method and choose, as a reference point, $z_0 = 1\,\textrm{GeV}^{-1}$. Then, substituting $\lambda(z)$ into the beta function and afterwards in the potential 
(\ref{potencialanomalous}), we solve numerically the corresponding $1d$-Schr\"odinger-like equation to find the first four glueball masses which depends on three parameters $c$, $\lambda_{\ast}$ and $\lambda_0$. 

\begin{figure}[!h]
\begin{center}
\includegraphics[width=6cm, height=4cm]{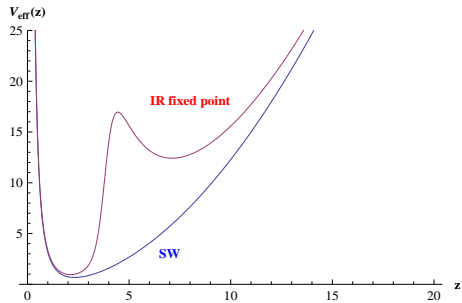}
\caption{\label{potentialsIRfixedpoint}The effective potentials from the minimal (\ref{potencial0}) and improved (\ref{potencialanomalous}) Soft-Wall models with the IR fixed point beta function (\ref{betalambdastar}). The values of the parameters are those listed in Table \ref{tablelambdastar}.}
\end{center}
\end{figure}

For the isotropic lattice results \cite{Meyer:2004gx} we are interested in, the best set of parameters able to fit reasonably well the first four masses is $c=-0.36\, \textrm{GeV}^2 $, $\lambda_{\ast}=350$ and $\lambda_0=18.5$. The corresponding masses are shown in Table \ref{tablelambdastar} while the effective potentials for the minimal and the improved Soft-Wall models are displayed in Figure \ref{potentialsIRfixedpoint}.  The mass ratios defined in Eq. (\ref{ratios}) take the values:
\begin{equation}
\label{ratiosBetaFunction1}
R_1\,=\,1.54 ; \quad R_2\,=\,2.04 ; \quad R_3\,=\, 2.42\,.
\end{equation}
\noindent The result for $R_1$ coincides with the result of the anisotropic lattice
and is out of the range of values from isotropic lattice. The values for $R_2$ and $R_3$ are fairly close to the range of isotropic lattice results, with respectively 10\% and 11\% of mismatch.

\begin{table}[!h]
\centering
\begin{center}
  \begin{tabular}{|c | c | c || c | c | c | c |}
    \hline\hline
     $c$ & $\lambda_0$ & $\lambda_{\ast}$ & $0^{++}$ & $0^{++\ast}$ & 
      $0^{++\ast\ast}$ & $0^{++\ast\ast\ast}$  \\ \hline
      -0.36 & 18.5 &  350  & 1.497 & 2.307 & 3.056 & 3.625 \\ \hline\hline 
  \end{tabular}
\end{center}
\caption{Scalar glueball masses from the phenomenological beta function with an IR fixed point at finite coupling (\ref{betalambdastar}).  
Masses and $c$ are expressed in GeV and GeV$^2$ respectively. $\lambda_0$ and $\lambda_{\ast}$ are dimensionless.}
\label{tablelambdastar}
\end{table}
%%%%%%%%%%%%%%%%%%%%%%%%%%%%%%%%%%%%%%%%%%%%%%%%%%%%%%%%%
%%%%%%%%%%% data from Frederic on 04-07-12 %%%%%%%%%%%%%%
%%%%%%%%%%%%%%%%%%%%%%%%%%%%%%%%%%%%%%%%%%%%%%%%%%%%%%%%%

 Our fit consistently predicts a coupling $\lambda_0$ at the scale $z_0=1$ GeV$^{-1}$ which is bigger\footnote{Note that the WA estimate was obtained from many various processes involving quark flavours ($\tau$ and heavy quarkonia  decays, lattice QCD, deep inelastic scattering, etc).} than the world average (WA) value  of the strong coupling \cite{Bethke:2009jm}: 
\begin{equation}\label{lambdazero}
\alpha_s=0.1184\;\;\Rightarrow\;\;\lambda_0^{(WA)}=4.464\;\;\;\;(N_c=3) \,\,,
\end{equation}
corresponding to the $Z^0$ boson mass $ M_{Z^0} = 91.2 {\rm GeV}\,$.

Interestingly, $c$ turns out to be negative with an absolute value about twice larger than the dilaton parameter $c=0.2325$ GeV$^{2}$ (\ref{SWc}) evaluated in the Soft-Wall model from the $\rho$ meson mass. A negative dilaton parameter has been introduced in the Soft-Wall models of QCD in \cite{brodsky:2010} and remains a subject of debate (see, \emph{e.g.} \cite{FJ,neg:2010}). On the other hand, a positive $c$ gives rise to a less satisfactory spectrum, most of the masses being outside their uncertainty intervals. A reason for that can be seen in Eq.(\ref{potencialanomalous}) where a positive $c$ increases the minimum of the effective potential.

As the first two excited states in Table 3 appear a bit too small, one could try to slightly vary the parameters. Letting the two other parameters $\lambda_{\ast}$ and $\lambda_0$ unchanged, to increase $c$ obviously makes the potential steeper in the infrared and increases indeed the masses. Then, it appears a lower bound for $c$ around $-(0.47)$ GeV$^2$ for which the ground-state mass reaches the upper bound allowed by \cite{Meyer:2004gx}. Nevertheless, the rest of the spectrum grows too slowly: while the second excited state becomes close to its lower bound ($3.112$ GeV compared with $3.120$ GeV), the first excited state remains too small ($2.369$ GeV compared with $2.565$ GeV). The fourth mass still remains within its uncertainty range. The same mechanism occurs if we decide instead to increase $\lambda_0$ with the upper bound $\lambda_0=19.4$. Finally, increasing $\lambda_{\ast}$ raises the peak shown in Figure \ref{potentialsIRfixedpoint} for intermediate $z$ without modifying its location. As a result, the potential takes for the lowest-lying states the shape of a well potential: for $\lambda_ {\ast}$ large enough, their masses become invariant and reach for the first three masses 1.502, 2.388 and 3.238 GeV respectively, the first excited state being as usual too light. A reason for this general fact is the important gap between the ground-state and the first excited state \cite{Meyer:2004gx} (around 1.3 GeV, to be compared, for instance, with the smaller gap of 0.94 GeV for anisotropic lattice \cite{Morningstar:1999rf}).

%%%%%%%%%%%%%%%%%%%%%%%%%%%%%

\subsubsection{Beta function with a linear IR asymptotic behavior}

\begin{figure}[!h]
\begin{center}
\includegraphics[width=6cm, height=4cm]{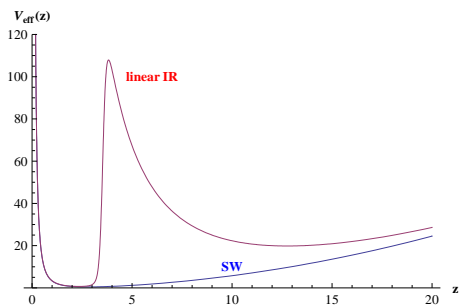}
\caption{\label{potentialslinearIR}The effective potentials from the minimal (\ref{potencial0}) and improved (\ref{potencialanomalous}) Soft-Wall models with the linear IR beta function (\ref{lambdalinear}). The values of the parameters are listed in Table \ref{tablelinear}.}
\end{center}
\end{figure}

The model beta function \cite{Zeng:2008sx,Ryttov:2007cx}: 
\begin{equation}\label{lambdalinear}
\beta(\lambda)=-\frac{b_0\lambda^2}{1+b_1\lambda} \qquad \qquad (b_0, b_1>0)
\end{equation}
behaves like the perturbative QCD beta function at 1-loop order 
and decreases asymptotically as $-\lambda$ in the infrared. Solving (\ref{betaf}) for this beta function, we find:
\begin{equation}
\lambda(z)=\frac{1}{b_1 W\left(\frac{e^{\frac{1}{b_1\lambda_0}}}{b_1\lambda_0}\left(\frac{z_0}{z}\right)^{b_0/b_1}\right)}\,\,.
\end{equation}
Then, the $5d$ potential reads as
\begin{equation}
V(z)=c^2z^2+\frac{15}{4z^2}+2c+\frac{1}{z^2}\frac{b_0}{b_1}
\frac{\Big[4\left(1+W(\frac{e^{\frac{1}{b_1\lambda_0}}}{b_1\lambda_0}(\frac{z_0}{z})^{b_0/b_1})\right)^2
+\frac{b_0}{b_1}\Big]}{\Big[1+W(\frac{e^{\frac{1}{b_1\lambda_0}}}{b_1\lambda_0}(\frac{z_0}{z})^{b_0/b_1})
\Big]^4}
\end{equation}
which presents the same subleading infrared and ultraviolet asymptotic behaviours than the beta function (\ref{betalambdastar}) considered before. Thus, the AdS effective potential gets modifications only for intermediate $z$. In particular, the squared masses $m_n^2$ still follows a linear Regge trajectory for large $n$. Figure \ref{potentialslinearIR} shows the effective potentials for the values of parameters listed in Table \ref{tablelinear}. 

With the beta function (\ref{lambdalinear}), the masses depend on the parameters $c$, $b_1$ and $\lambda_0$. The result of the fit of the first four masses is shown in Table 4 and corresponds to the set of parameters $c=-0.25$ GeV$^2$, $b_1=1.2\times 10^{-3}$ and $\lambda_0=19$. 
The mass ratios defined in Eq. (\ref{ratios}) now assume the values:
\begin{equation}
\label{ratiosBetaFunction2}
R_1\,=\,1.64 ; \quad R_2\,=\,2.25 ; \quad R_3\,=\, 2.86\,.
\end{equation}
\noindent The result for $R_1$ is within the error bar for the result of the anisotropic lattice
and is out of the range of values from isotropic lattice. The values for $R_2$ and $R_3$ are 
inside the range of isotropic lattice results.

In this case, the dilaton parameter is negative and, furthermore, very close in absolute value to the minimal Soft-Wall estimate (\ref{SWc}). The coupling $\lambda_0$ we figured out, fitted at the energy scale $z_0=1$ GeV$^{-1}$, is quite close to the $\lambda_0$ we found in the case of the IR fixed point beta function (19 compared with 18.5), showing consistency in our results.  

\begin{table}[!h]
\centering
\begin{center}
  \begin{tabular}{|c | c | c || c | c | c | c | c |}
    \hline\hline
     $c$ & $b_1$ & $\lambda_0$ & $0^{++}$ & $0^{++\ast}$ & 
      $0^{++\ast\ast}$ & $0^{++\ast\ast\ast}$  \\ \hline
     -0.25 & $1.2\times 10^{-3}$ & 19 & 1.484 & 2.434 & 3.346 & 4.239 \\ \hline\hline 
  \end{tabular}
\end{center}
\caption{Glueball masses in GeV for the the beta function (\ref{lambdalinear}).  
$c$ is expressed in GeV$^2$ while $b_1$ and $\lambda_0$ are dimensionless.}
\label{tablelinear}
\end{table}

Changing one parameter while keeping fixed the other two allows one to see that, on the one hand, $c$ and $\lambda_0$ are bounded by the same limit values as derived for the IR fixed point beta function. On the other hand, the parameter $b_1$ plays the same role than the fixed point $\lambda_{\ast}$ such that, for $b_1$ large enough, the lowest-lying masses become independent of $b_1$\footnote{While $b_1$ in (\ref{lambdalinear}) has nothing to do with $b_{1}^{(pert)}$ of the perturbative QCD beta function, let's remark that they seem numericaly close ($b_1^{(pert)}\approx0.9\times 10^{-3}$ for $n_f=0$). Nevertheless, the shape of the $5d$ potential displayed in Figure \ref{potentialslinearIR} is very sensitive to the value of $b_1$.}. 

From our analysis, it thus appears that the model beta function (\ref{lambdalinear}) seems favoured with respect to the previous case (\ref{betalambdastar}). Indeed, the former allows one to fit the first four masses reasonably well with the exception only of the first excited state whose the gap with the ground-state might be, as discussed at the end of subsection {\bf 3.2.1}, too high regarding other studies \cite{Morningstar:1999rf}.

%While the beta function (\ref{lambdalinear}) cannot reproduce the perturbative QCD beta function at 2-loop order, we can reduce the number of parameters fixing $b_1$ to its perturbative value when $n_f=0$, namely $b_1=\frac{51}{121}b_0^2=9.10^{-4}$. Then, the paramters  $c=-0.2$ and $\lambda_0=18.67$ and the masses are:\\
%\begin{table}[!h]
%\centering
%\begin{center}
 % \begin{tabular}{|c | c || c | c | c | c | c |}
  %  \hline\hline
   %  $\sqrt{|c|}$ & $\lambda_0$ & $0^{++}$ & $0^{++\ast}$ & 
    %  $0^{++\ast\ast}$ & $0^{++\ast\ast\ast}$  \\ \hline
     %0.45 & 18.67 & 1.468 & 2.428 & 3.351 & 4.258 \\ \hline\hline 
  %\end{tabular}
%\end{center}
%\caption{Glueball masses in GeV for the the beta function (\ref{lambdalinear}).  
%$\sqrt{|c|}$ is expressed in GeV whereas $\lambda_0$ is dimensionless.}
%\end{table}

%%%%%%%%%%%%%%%%%%%%%%%%%%%%%%%%%%%%%%%%%%%%%%%%%%%%

\subsubsection{Beta function with a cubic IR asymptotic behavior}\label{section3.1}
 
We consider now the following beta function \cite{Zeng:2008sx}:
\begin{equation}\label{cubicIRbehavior}
\beta(\lambda)=-b_0\lambda^2-b_1\lambda^3 \qquad \qquad (b_0\;, \; b_1>0)\;
\end{equation}
which behaves as $-\lambda^3$ in the infrared.

The solution of the Renormalization Group equation reads
\begin{equation}\label{renor}
\lambda(z)=-\frac{b_0}{b_1}\frac{1}{1+W_{-1}\left(-\left(\frac{z}{z_0}\right)^{\frac{b_0^2}{b_1}}\left(1+\frac{b_0}{b_1\lambda_0}\right)e^{-(1+\frac{b_0}{b_1\lambda_0})}\right)}
\end{equation}
%thus 
%\begin{equation}
%\lambda(z)=\frac{1}{b_0\ln(\frac{z_0}{z})+\frac{b_1}{b_0}\ln(\frac{b_1\lambda+b_0}{b_1\lambda_0+b_0}
%\frac{\lambda_0}{\lambda})}
%\end{equation}
imposing $\lambda(z_0)=\lambda_0$. $W_{-1}(x)$ is the generalized Lambert function of order -1 which leads, on the one hand, to the QCD-like perturbative behaviour (\ref{QCDlike}) at small coupling. On the other hand, because of the analytical properties of $W_{-1}(x)$, the coupling $\lambda$ becomes infinite for a finite value of $z$ given by
\begin{equation}
{z}_{max}
=\frac{z_0 e^{\frac{1}{b_0\lambda_0}}}{\left(1+\frac{b_0}{b_1\lambda_0}\right)^{b_1/b_0^2}}\,\,.
\end{equation}
In other words, the implementation of the beta function (\ref{cubicIRbehavior}) into the Soft-Wall framework gives rise to an AdS slice $0< z < z_{max}$. In contrast with the Soft-Wall model, the $5d$ potential (\ref{potencialanomalous})
\begin{equation}\label{potzmax}
V(z) =c^2z^2+\frac{15}{4z^2}+2c-\frac{1}{z^2}b_1 \lambda(z)^2(4-b_1\lambda(z)^2)\,\,
\end{equation}
goes to infinity at $ z = z_{max}$: unexpectedly, there is the formation 
of a hard wall located at $z_{max}$ (independent of $c$) and the solutions of the Schr\"odinger-like equation (\ref{Schroe}) will be non-vanishing only in the region $0 < z < z_{max}$. 

\begin{figure}[!h]
\begin{center}
\includegraphics[width=7cm, height=5cm]{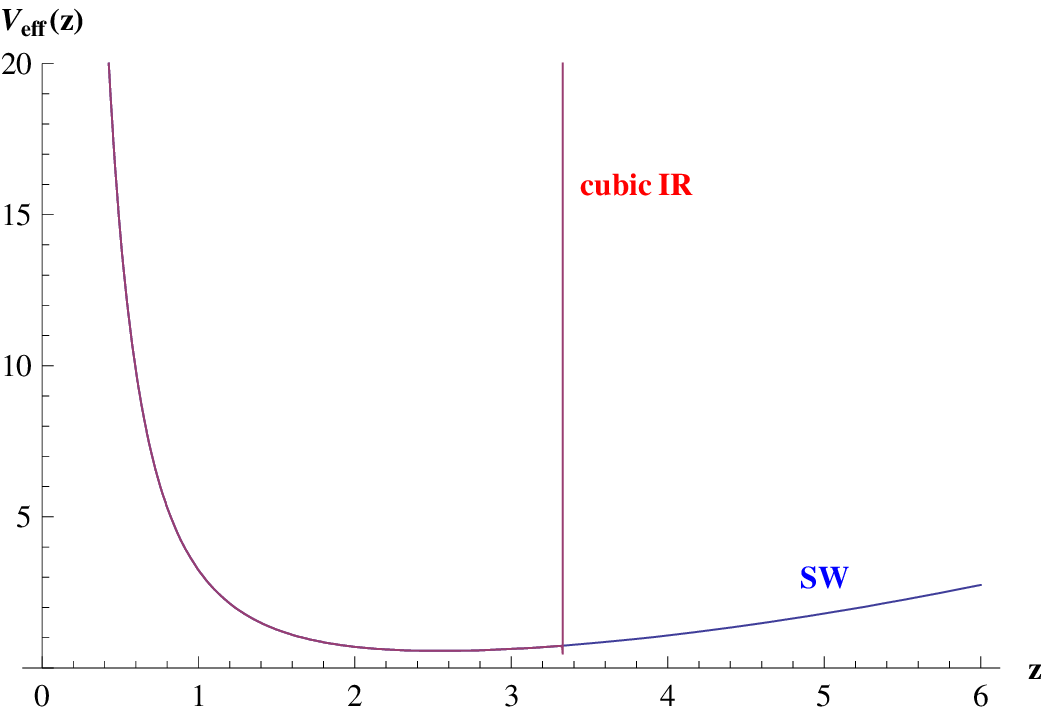}
\includegraphics[width=7cm, height=5cm]{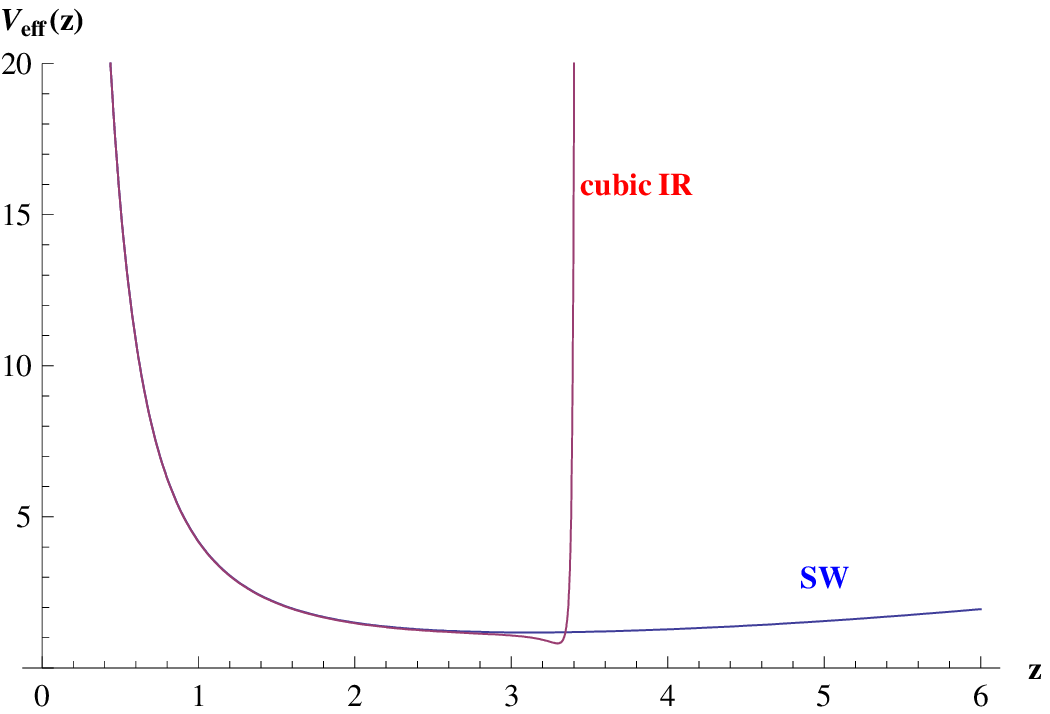}
\caption{\label{figureCubicIR}The effective potentials from the minimal and improved Soft-Wall models with the cubic IR beta function (\ref{cubicIRbehavior}) for $c$ negative (left panel) and $c$ positive (right panel). The values of the parameters are listed in Table \ref{tabzmax}.}
\end{center}
\end{figure}

The fact that the background was initially of the Soft-Wall model does not have significant influence 
in getting the glueball mass spectrum in this case, 
as shown in Table \ref{tabzmax} and Figure \ref{figureCubicIR} where a positive as well as negative dilaton 
parameter $c$ is in fact an acceptable value. Furthermore, the $5d$ potential having the shape of a sharp well, the Regge-like behaviour of the squared masses is lost and we have instead $m_n^2\sim n^2$ for large $n$.
\begin{table}[!h]
\begin{center}
  \begin{tabular}{| c | c | c | c || c | c | c | c |}
    \hline\hline
 $c $ & $ b_1 $ & $ \lambda_0$ & ${z}_{max}$ & $0^{++}$  
   & $0^{++\ast}$ & $0^{++\ast\ast}$ & $0^{++\ast\ast}$ \\ \hline
 -0.3 & $10^{-9}$ & 17.9 & 3.330 & 1.476 & 2.480 & 3.453 & 4.415 \\ \hline
 0.05 & $1.3\times10^{-5}$ & 17 & 3.436 & 1.522 & 2.465 & 3.393 & 4.315 \\ \hline
            \hline
  \end{tabular}
\end{center}\caption{Masses for the scalar glueball ground-state and the radial excitations $J^{PC}=0^{++}$
for the phenomenological beta function (\ref{cubicIRbehavior}) with a cubic IR asymptotic behavior.  Masses are expressed in GeV, $c$ in GeV$^2$ and $b_1$ is dimensionless. For completeness, we also show the values of $z_{max}$ expressed in GeV$^{-1}$.}
\label{tabzmax}
\end{table}

  The mass ratios for the two lines of Table \ref{tabzmax} take, respectively,  the values:
\begin{eqnarray}
\label{ratiosBetaFunction3}
& & R_1\,=\,1.68 ; \quad R_2\,=\,2.34 ; \quad R_3\,=\, 2.99\,\cr
& & R_1\,=\,1.62 ; \quad R_2\,=\,2.23 ; \quad R_3\,=\, 2.84\,\
\end{eqnarray}
\noindent  The results for $R_1$ are within the error bar for the result of the anisotropic lattice
and out of the range of values from isotropic lattice. The values for $R_2$ are 
inside the range of isotropic lattice results. For $R_3$ the first line presents a mismatch of 10\% with respect to the isotropic lattice while the second line is inside the error bar.

  Both for a positive and a negative dilaton parameter, the first excited state is slightly too light although the result seems a bit better than for the IR linear and the IR fixed point beta functions considered in the subsection {\bf 3.2.1} and {\bf 3.2.2}. Nevertheless, the fourth mass is either too large (for $c$ negative) or approaches closely its upper bound (for $c$ positive). This result was expected since the masses of the higher states are significantly increased by the presence of the hard wall.

%%%%%%%%%%%%%%%%%%%%%%%%%%%%%%%%%%%%%%%%%%%%%%%%%%%%%%%%%
%%%%%%%%%%% data from Frederic on 03-07-12 %%%%%%%%%%%%%%
%%%%%%%%%%%%%%%%%%%%%%%%%%%%%%%%%%%%%%%%%%%%%%%%%%%%%%%%%

%%%%%%%%%%%%%%%%%%%%%%%%%%%%%%%%%%%%%%%%%%%%%%%%%%%

\section{Phenomenological beta functions from Regge-like glueball spectroscopy}

The glueball masses obtained from lattice studies shown in Table 1  lie approximately in  
linear Regge trajectories, namely, the square of the masses has approximately a linear relation 
with the radial quantum number. In this section, we build up beta functions that reproduce 
this behaviour. In particular, we will consider the four masses from the isotropic lattice 
calculations shown in the second column of Table 1. A linear Regge trajectory which fits these masses corrresponds to
\begin{eqnarray}
m_n^2 &=& 4.50 n   + 2.51  \,, \label{linearfit}
\end{eqnarray}
where the squared masses are given in ${\rm GeV}^2$.
This Regge trajectory implies the glueball masses given in Table \ref{ajustemassas}.
The first mass is just 1$\%$ above the lattice result including the error bar. The other three
masses are within the error bars.

\begin{table}[!h]
\centering
\begin{center}
  \begin{tabular}{| c | c | c | c |}
    \hline\hline
      $0^{++}$ & $0^{++\ast}$ & $0^{++\ast\ast}$ & $0^{++\ast\ast\ast}$  \\ \hline\hline
     1.585 & 2.648 & 3.393 & 4.001 \\ \hline\hline 
  \end{tabular}
\end{center}
\caption{Glueball masses in GeV from the Regge trajectory of Eq.(\ref{linearfit}).}
\label{ajustemassas}
\end{table}

In order to obtain the classes of beta functions that lead to this Regge linear spectrum, we consider 
the Soft-Wall potential modified by the contribution from the anomalous dimension
of the scalar glueball operator (\ref{potencialanomalous}):
\begin{eqnarray}\label{potencialanomalous2}
V(z)
&=& c^2z^2+\frac{15}{4z^2}+2c+\frac{1}{z^2}\left[4+f(z) \right] f(z) 
\end{eqnarray}
where  
\begin{eqnarray}\label{f(z)}
f(z)\equiv \frac{d\beta(\lambda)}{d\lambda}-\frac{2}{\lambda}\beta(\lambda) \,.
\end{eqnarray}

It is convenient to rewrite $f(z)$ in terms of the coupling $\lambda(z)$:
\begin{eqnarray}\label{fdef}
f(z) &=& -1 - z\left( \frac{\lambda''}{\lambda'}-2\frac{\lambda'}{\lambda}\right) \cr
&=&-1 -z \frac{d}{dz}\left( \ln \frac{\lambda'}{\lambda^2} \right)
\end{eqnarray}
where a prime denotes a derivative with respect to $z$.

We use the following ansatz:
\begin{eqnarray}\label{ansatz}
(f(z)+2)^2 = k +k_1 z^2 + k_2 z^4\,,
\end{eqnarray}
with $k$, $k_1$ and $k_2$ constants, that will be fixed in terms of $c$, as we show in the sequel. 
The positivity of this ansatz requires:
\begin{equation}\label{positivity}
 k_1^2 \le 4kk_2 \,.
\end{equation}
From the Eqs.(\ref{fdef}) and (\ref{ansatz}), we have thus to solve the differential equation:
\begin{equation}\label{eqsign}
\frac{d}{dz}\ln\left(\frac{\lambda'(z)}{\lambda^2(z)}\right)
=\frac{1}{z}\mp\frac{1}{z}\sqrt{k +k_1z^2+k_2z^4}\,.
\end{equation}

To reproduce the QCD asymptotic behaviour  at leading order shown in Eq.(\ref{QCDlike}), 
we impose that
\begin{equation}\label{zto0b0}
\lim_{z\to 0}\left(\frac{1}{\lambda(z)}\right)'=-\frac{b_0}{z}\;,
\end{equation}
which, in turn, implies:
\begin{equation}
\lim_{z\to 0} \frac{d}{dz}\ln\left(\frac{\lambda'(z)}{\lambda^2(z)}\right)
= - \frac{1}{z}\,.
\end{equation}
Hence, we must have $k=4$. For small $z$, we also must take the upper sign in Eq.(\ref{eqsign}). Note that this analysis of the perturbative small $z$ limit does not impose the sign of Eq.(\ref{eqsign}) for large values of $z$ as it will be discussed below.

Then, the potential takes the form:
\begin{equation}
V(z)=\left(c^2+k_2\right) z^2+\frac{15}{4z^2}+\left(2c+k_1\right)\,.
\end{equation}
Solving the $1d$-Schr\"odinger-like equation (\ref{Schroe}) with this potential, one finds the following $4d$ mass spectrum:
\begin{equation}\label{spectrumk1k2}
m_n^2=4\sqrt{c^2+k_2}\,n+6\sqrt{c^2+k_2}+2c+k_1 \,.
\end{equation}
This equation is matched to the glueball linear Regge trajectory of Eq.(\ref{linearfit}) yielding:
\begin{equation}
\left\{
\begin{array}{ccc}\label{matching2}
k_1(c)&=&-\frac{106}{25}\, - 2c\,\,,\\
k_2(c)&=&\frac{81}{64}\, - c^2\,\,.
\end{array}
\right.
\end{equation}
With the positivity condition (\ref{positivity}), the allowed range of the remaining parameter $c$ is $c_{min}\le c \le c_{max}$ with
\begin{eqnarray}\label{cmincmax}
c_{min}=  - (0.983\, {\rm GeV})^2 \qquad \textrm{and} \qquad c_{max} = (0.344\, {\rm GeV})^2\,.
\end{eqnarray}
The extreme values of the dilaton parameter saturate the positivity condition, \emph{i.e.}, $k_1^2=16k_2$. Also here, we observe the possibility of negative values for $c$ and the fact that the allowed range of negative values is almost ten times larger than the allowed range for positive values of $c$. We also note that $k_1$, expressed in terms of $c$, can only be negative. 

Let us now determine the corresponding beta functions. They will depend on the two parameters $c$ and $\lambda_0$. We will treat separately the cases $k_1^2 = 16k_2$ (extreme case) and $k_1^2 < 16k_2$ (non-extreme case) since then the differential equation (\ref{eqsign}) assumes different forms.

\subsection{Extreme case}

In this case, $c$ is fixed and can take the two values $c_{min}$ and $c_{max}$, corresponding to $k_1=-2.307$ and $k_1=-4.477$ respectively. Then, Eq.(\ref{eqsign}) reduces to:
\be
 1 - z \frac{d}{dz}\left( \ln \frac{\lambda'}{\lambda^2} \right)= \pm (2+\frac{k_1}{4}z^2)\,\,.
 \label{2diff equations}
\ee
\noindent 
Since $k_1$  is negative, there is a critical value of $z$ given by $z_c^2= - {8}/{k_1}$. We get $z_c=1.862$ GeV$^{-1}$ and $z_c=1.337$ GeV$^{-1}$ for $c=c_{min}$ and $c=c_{max}$ respectively. For $z < z_c$, the sign of Eq.(\ref{2diff equations}) is fixed by the perturbative limit. For $z \ge z_c$, we are free to choose the sign without spoiling the smoothness of the coupling $\lambda (z)$. 
In the following, we are going to consider the two possibilities.

$\bullet$ First, we take the upper sign of Eq.(\ref{2diff equations}) for all values of $z$ and 
find an analytical solution:
\begin{equation}
\lambda(z)=\frac{\lambda_0}{1-b_0\frac{\lambda_0}{2}
\left[Ei(-\frac{k_1}{8}z^2)-Ei(-\frac{k_1}{8}z_0^2)\right]}\,\,,
\label{lambdalimitingcases}
\end{equation}
expressed in terms of the exponential integral  function $Ei(x)$. Furthermore, we will choose to fix the remaining parameter $\lambda(z_0)=\lambda_0$ at the energy scale $z_0=1/M_{Z^0}$ given by the 
mass of $Z^0$ and which corresponds to the world average value of the 't Hooft coupling $\lambda^{(WA)}=4.464$ (\ref{lambdazero}). While $\lambda_0$ with $n_f=0$ is not equal to $\lambda^{(WA)}$, it is not very different in the perturbative regime. By slightly varying $\lambda_0$ around its WA value in Eq.(\ref{lambdalimitingcases}), we will observe no strong effect in the behaviour of the coupling constant. On the contrary, we will see in the  subsection {\bf 4.2} when $c$ is not extreme that a small variation of $\lambda_0$ can drastically change the low energy behaviour of the coupling. 

The beta function reads
\begin{equation}
\beta(z)\equiv-z\lambda'(z)=-\frac{\lambda_0^2 b_0e^{-\frac{k_1}{8}z^2}}
{\Big\{1-b_0\frac{\lambda_0}{2}\left[Ei(-\frac{k_1}{8}z^2)-Ei(-\frac{k_1}{8}z_0^2)\right]\Big\}^2}\;.
\label{betalimitingcases}
\end{equation}

\begin{figure}[!h]
\begin{center}
\includegraphics[width=6cm, height=4cm]{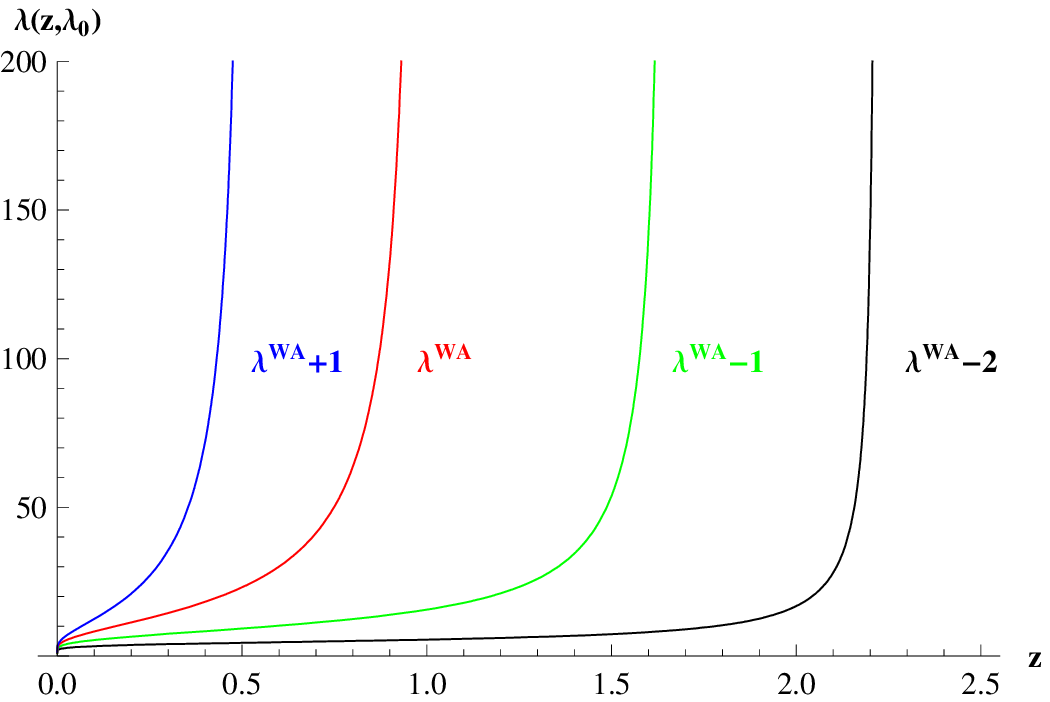}
\includegraphics[width=6cm, height=4cm]{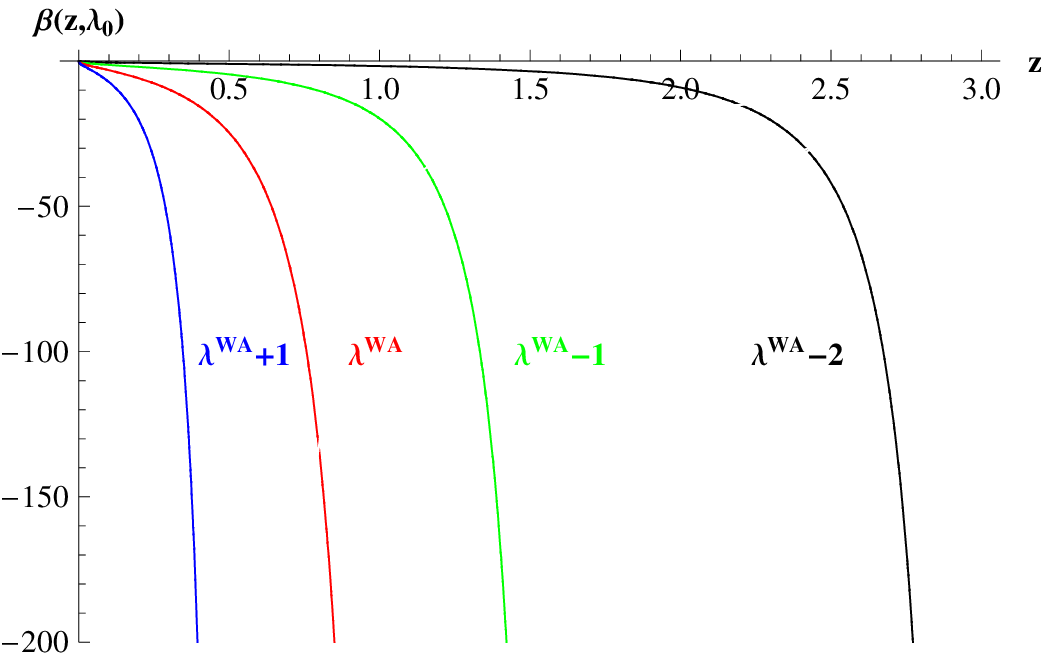}
\includegraphics[width=6cm, height=4cm]{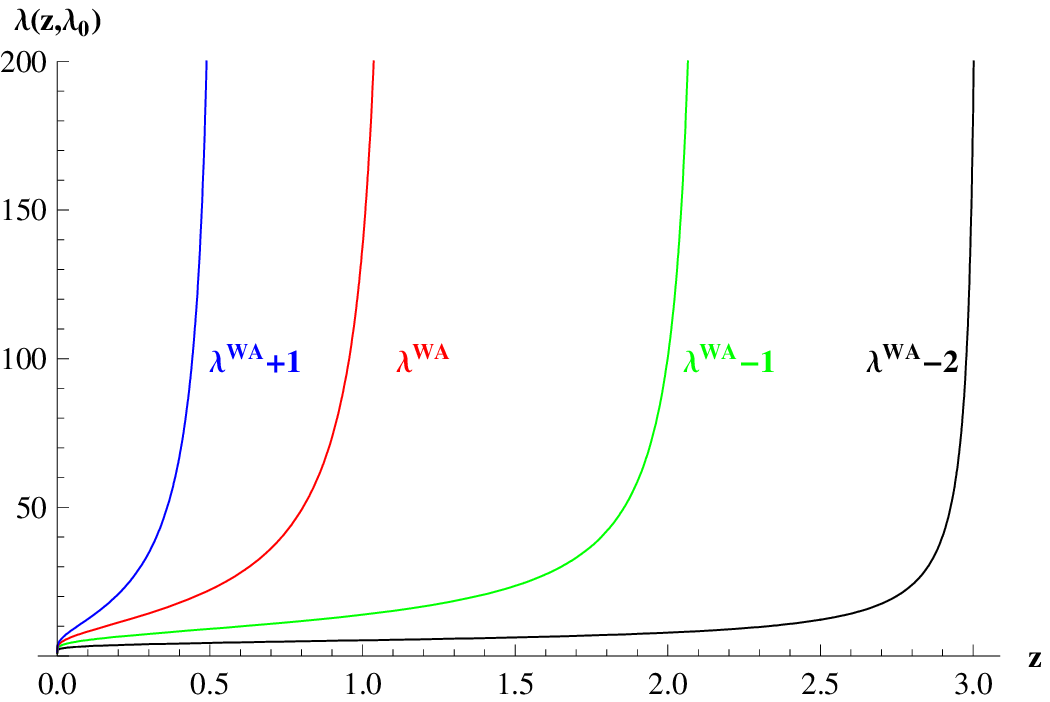}
\includegraphics[width=6cm, height=4cm]{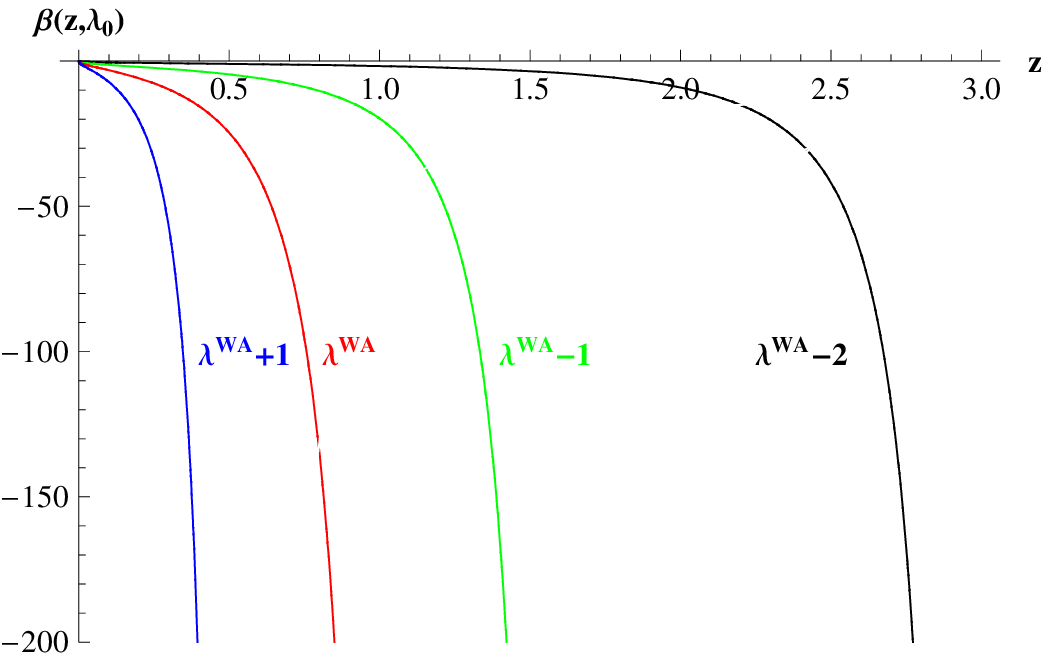}
\caption{\label{figure1} The 't Hooft couplings $\lambda(z)$ 
and the phenomenological beta functions $\beta(z)$ as defined in (\ref{lambdalimitingcases}) and (\ref{betalimitingcases})
for different values of $\lambda_0$ close to $\lambda_0^{(WA)}$ and $c=c_{max}$ (upper panels) and
$c=c_{min}$ (lower panels).}
\end{center}
\end{figure}

\begin{figure}[!h]
\begin{center}
\includegraphics[width=7cm, height=5cm]{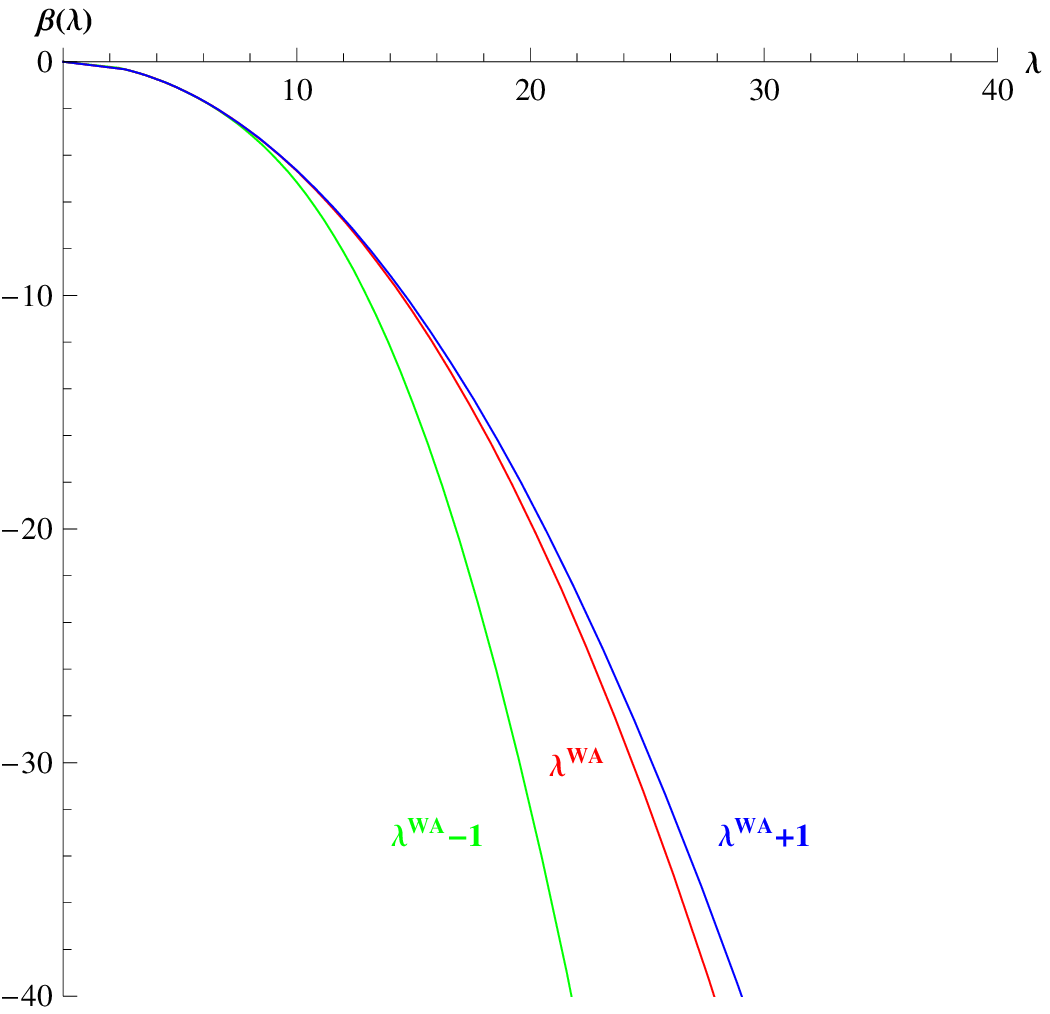}
\includegraphics[width=7cm, height=5cm]{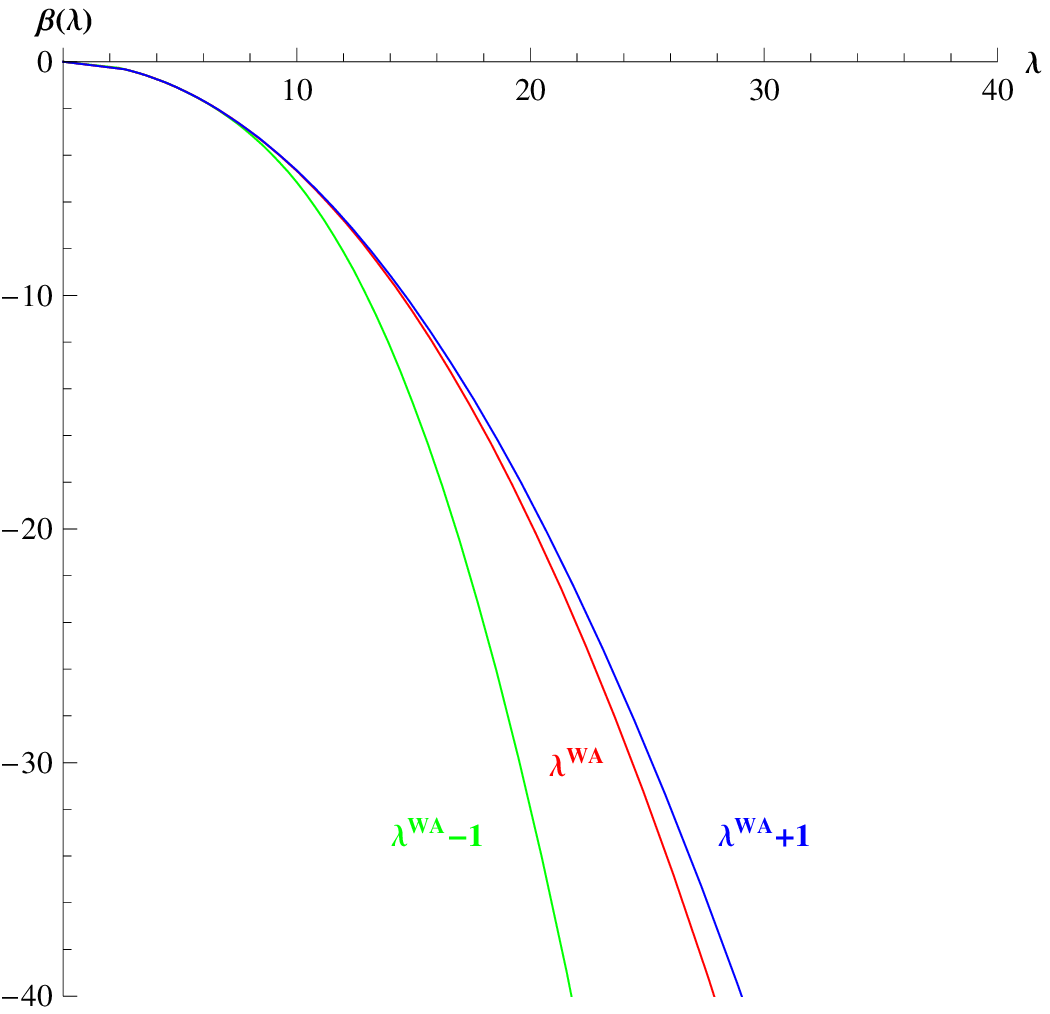}
\caption{\label{figure2} The phenomenological beta functions 
(\ref{betalimitingcases}) in terms of $\lambda$ for different values of $\lambda_0$ close to $\lambda_0^{(WA)}$ when $c=c_{max}$ (left panel) and $c=c_{min}$ (right panel).}
\end{center}
\end{figure}

Considering the two possible values for $c$ which fixes $k_1$ and $k_2$, we have plotted in Figure \ref{figure1} the couplings and the beta functions in terms of the holographic coordinate 
$z$ for four values of $\lambda_0$ around $\lambda^{(WA)}$. 
Both $\lambda(z)$ and $\beta(z)$ diverge at a finite value of $z=z_{max}$
where the denominators in the Eqs.(\ref{lambdalimitingcases}) and (\ref{betalimitingcases}) vanish. $z_{max}$ depends on $\lambda_0$ and is larger for $c_{min}$ than for $c_{max}$. 
In Figure \ref{figure2}, we show the dependence of the beta functions
on the coupling $\lambda$ for the two extreme values of $c$. 

$\bullet$ Secondly, when we choose for the Eq.(\ref{2diff equations}) the following form: 
\begin{equation}\label{extremecasebis}
 z \frac{d}{dz}\left( \ln \frac{\lambda'}{\lambda^2} \right)= \left\{ 
\begin{array}{cc}
 -(1+\frac{k_1}{4}z^2)\;\; \quad &   z < z_c\,\,, \\
(3+\frac{k_1z^2}{4})\;\; \quad &  z\geq z_c\,\,,
\end{array}
\right.
\end{equation}
%The solution is  
%where $\Theta(z)$ is the step function and $\lambda_{<}(z)$ and $\beta_{<}(z)$ are the analytical solutions (\ref{lambdalimitingcases}) and (\ref{betalimitingcases}) of the previous subsection. 
with a change of sign at $z_c$, we observe that the shapes of the 't Hooft coupling and of the beta function are similar to the previous case, shown in Figure \ref{figure1}, and diverge at $z=z_{max}$. 

Depending on $\lambda_0$, $z_{max}$ can be smaller or larger than $z_c$. Note that decreasing $\lambda_0$ increases $z_{max}$. 
When $z_{max}<z_c$, only the solution (\ref{lambdalimitingcases}) is present. This happens when $\lambda_0=3.942$ (negative $c$) and $\lambda_0=3.712$ (positive $c$). When $z_{max}>z_c$, we have another solution, given by the complete solution of Eq.(\ref{extremecasebis}):
\be
\lambda(z)= \left(1-\Theta(z-z_c)\right)\lambda_{<}(z) +\frac{\Theta(z-z_c)}{\frac{1}{\lambda_{<}(z_c)}+\frac{b_0}{2}\left(\frac{z^2}5{z_c^2}+1\right)e^{2-\frac{z^2}{z_c^2}}-b_0e}
\ee
and
\be
\beta(z)=\left(1-\Theta(z-z_c)\right)\beta_{<}(z)-\frac{\Theta(z-z_c)}{\Big[\frac{1}{\lambda_{<}(z_c)}+\frac{b_0}{2}\left(\frac{z^2}{z_c^2}+1\right)e^{2-\frac{z^2}{z_c^2}}-b_0e\Big]^2}\frac{b_0 z^4}{z_c^4}e^{2-\frac{z^2}{z_c^2}}\,.
\ee
where $\Theta(z)$ is the step function and $\lambda_{<}(z)$ and $\beta_{<}(z)$ are given by (\ref{lambdalimitingcases}) and (\ref{betalimitingcases}).

Let us emphasize that changing the sign in Eq.(\ref{2diff equations}) leads to two different solutions. Nevertheless, the data we use do not allow us to select a unique phenomenological beta function. 
%Below, we display in Figure \ref{figurebetaoflambdaextreme} the dependence on $\lambda$ of the beta function for three values of $\lambda_0$ and the two possible extreme values of $c$. 

%\begin{figure}[!h]
%\begin{center}
%\includegraphics [width=6cm, height=4cm]{cMAX-ChangeOfSign-BetaAsFunctionOfLambda.eps}
%\includegraphics [width=6cm, height=4cm]{cMIN-ChangeOfSign-BetaAsFunctionOfLambda.eps}
%\caption{\label{figurebetaoflambdaextreme} The phenomenological beta functions $\beta(\lambda)$ as a function of $\lambda$ for different values %of $\lambda_0$ and the two possible values $c_{max}$ (left) and
%$c_{min}$ (right).}
%\end{center}
%\end{figure}

\subsection{Non-extreme case}

In this case, the square root on the \emph{r.h.s.} of Eq.(\ref{eqsign}) does not vanish for any value of $z$. Therefore, the upper sign is the only possibility here. Then, we find out:
\begin{eqnarray}
\left(\frac{1}{\lambda}\right)'=-\frac{b_0}{4z}
\frac{\left(\sqrt{4+{k_1}z^2+{k_2}z^4}+\frac{k_1}{4}z^2+2\right)
\left({k_1}+4\sqrt{k_2}\right)^{\frac{k_1}{4\sqrt{k_2}}}}
{\left(2\sqrt{k_2}\sqrt{4+{k_1}z^2+{k_2}z^4}+2{k_2}z^2
+{k_1}\right)^{\frac{k_1}{4\sqrt{k_2}}}}
e^{1-\frac 12 \sqrt{4+{k_1}z^2+{k_2}z^4}}\nonumber\\
\label{condition}
\end{eqnarray}
which cannot be solved analytically. The parameters are $\lambda_0$ and $c$. The latter, calculated from (\ref{positivity}) and (\ref{matching2}), can assume any value inside the range $ c_{min} < c < c_{max}$ (\ref{cmincmax}).

In order to analyse the behaviour of the general equation (\ref{condition}), 
let's write its integral from $z$ to $z_0$ under the form: 
\begin{equation}
\frac {1}{\lambda_0} - \frac{1}{\lambda(z)}= F(z,z_0)\;.\label{general F} 
\end{equation} 
Then, the 't Hooft coupling reads
\begin{equation}
\lambda(z)=\frac{\lambda_0}{1 - \lambda_0 F(z,z_0)}\;.\label{general solution}
\end{equation} 

For large enough values of $\lambda_0$, namely $\lambda_0>\lambda_0^{(limit)}=3.234$, there is always a $z=z_{max}$ at which $\lambda(z)$ diverges. Moreover, this occurs for any value of $c$. 
Then, the shapes of $\lambda(z)$, $\beta(z)$ and $\beta(\lambda)$ are similar to the ones displayed in Figures \ref{figure1} and \ref{figure2}.

On the contrary, for $\lambda_0<\lambda_0^{(limit)}$, the results turn out to be drastically different. 
In the sequel, we will consider the illustrative case $\lambda_0=3.164$. While seemingly totally arbitrary, this choice will allows us to discuss qualitatively all the possible behaviours the coupling and the beta function can take. We could have also chosen any other value for $\lambda_0$, the crucial point being that $\lambda_0$ must be smaller than $\lambda_0^{(limit)}$.

First, we observe that for $c$ close to the extreme values of its allowed range, the coupling solution 
remains singular. Like the two cases considered above, there is a value $z=z_{max}$ 
for which the denominator  $1-\lambda_0 F(z_{max},z_0)$ vanishes. 

\begin{figure}[!h]
\begin{center}
\includegraphics [width=6cm, height=4cm]{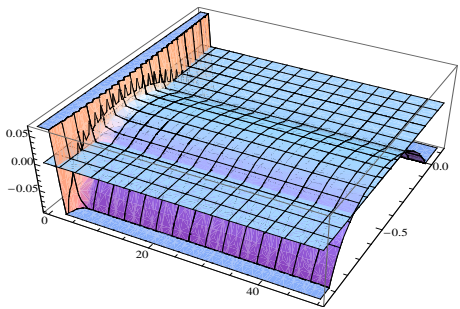}
\caption{\label{singularbehaviournf6} The function $\lambda_0/\lambda(z)$ 
and the zero-value plane surface. The coordinate $z$ and the parameter 
$c$ run respectively from 0 to 50 and from its extreme values 
$c_{min} \simeq -(0.983\,\textrm{GeV})^2$ to $c_{max} \simeq (0.344\,\textrm{GeV})^2$.}
\end{center}
\end{figure}

However, there is an interval for $c$, namely 
\begin{equation}
c\in ]-(0.860\;\textrm{GeV})^2,-(0.548\textrm{GeV})^2[\label{interval}
\end{equation}
for which this denominator never vanishes. Note that this interval only contains negative values of $c$. As a result, the coupling and the beta function 
are finite for all $z$. We show in Figure \ref{singularbehaviournf6} the 
behaviour of $1-\lambda_0 F(z,z_0)=\lambda_0/\lambda(z)$
as a function of $z$ and $c$. 

\begin{figure}[!h]
\begin{center}
\includegraphics [width=6cm, height=4cm]{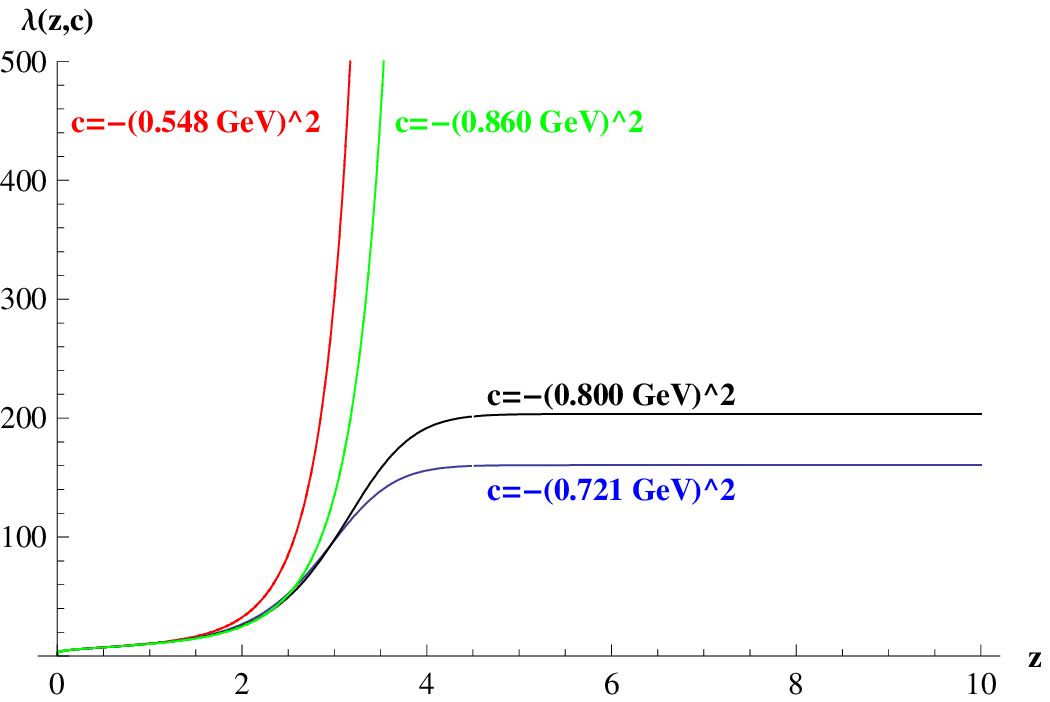}
\includegraphics [width=6cm, height=4cm]{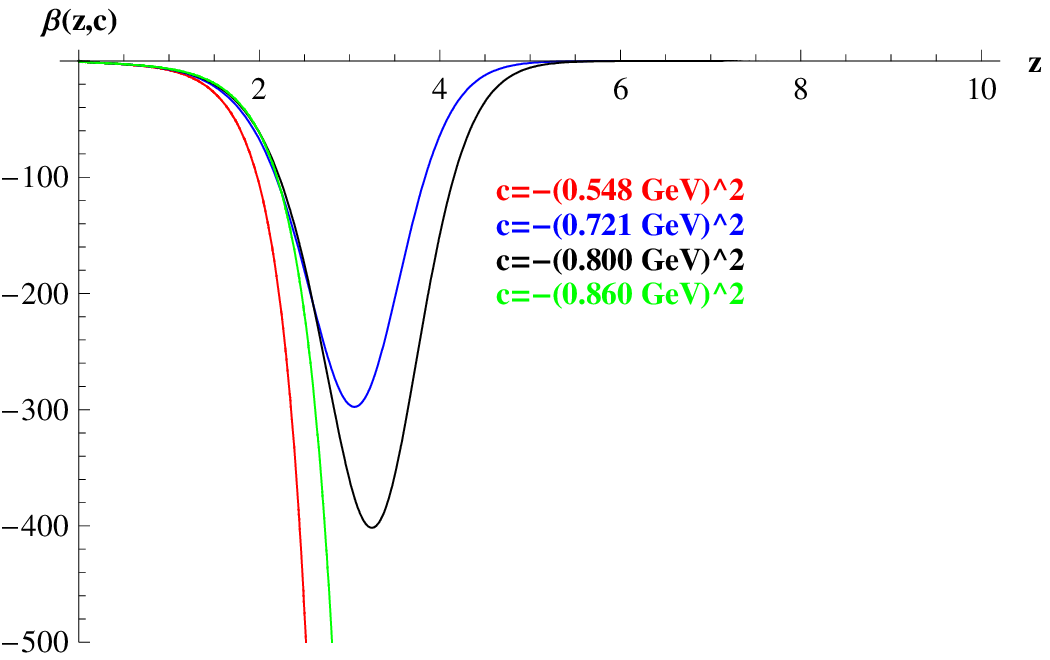}
\caption{\label{lambdanf6} The 't Hooft couplings $\lambda(z,c)$ and the associated 
beta function $\beta(z,c)$ in terms of $z$ for some remarkable values of $c$ in the case $\lambda_0<\lambda^{(limit)}$.}
\end{center}
\end{figure}

Further, in Figure \ref{lambdanf6}, we show the 't Hooft couplings and the phenomenological beta 
functions for some noticeable values of $c$. The red and green curves correspond 
to the two limit values $c=-(0.860\;\textrm{GeV})^2$ and $c=-(0.548\;\textrm{GeV})^2$ beyond which $\lambda(z)$ and 
$\beta(z)$ are singular at $z_{max}$. On the other hand, 
for any value of the parameter $c$ inside the interval (\ref{interval}), 
the 't Hooft coupling presents an \emph{IR fixed point} $\lambda_{\ast}$.
Accordingly, the beta function \emph{vanishes} for sufficiently large $z$. 
Figure \ref{lambdanf6} displays the case $c=-(0.800\;\textrm{GeV})^2$ and the 
mid-value $c=-(0.721\;\textrm{GeV})^2$ as well.
  It is important to stress that outside the interval  $ \,-(0.860\;\textrm{GeV})^2 < c < -(0.548\;\textrm{GeV})^2$ there is an infrared cut off represented by the maximum value of the fifth dimension: $z_{max}$ so that the model behaves as a hard wall and the Regge trajectories are not asymptotically linear outside that interval.  

\begin{table}[!h]
\centering
\begin{center}
  \begin{tabular}{| c | c | c |}
    \hline
    $c$ (GeV$^2$) & $z_{max}$ (GeV) & $\lambda_{\ast}$ \\ \hline\hline
    $-(0.548)^2$ & $4.246$ & \\ \hline
    $-(0.549)^2$ &  & 62314\\ \hline
    $-(0.721)^2     $ & & 161\\ \hline
    $-(0.800)^2$ & & 203\\ \hline
$-(0.859)^2$ & & 49819\\ \hline
$-(0.860)^2$ & 4.496 &\\ \hline
  \end{tabular}
\end{center}
\caption{Numerical estimates for $z_{max}$ and $\lambda_{\ast}$ for some values of the 
dilaton parameter inside the interval (\ref{interval}) and immediately beyond.}
\label{intermediatecnf6}
\end{table}

In Table \ref{intermediatecnf6}, we list the numerical estimates for $z_{max}$ 
and $\lambda_{\ast}$ for some characteristic values of $c$. Note that $\lambda_{\ast}$ takes very different values. 

\begin{figure}[!h]
\begin{center}
\includegraphics [width=8cm,height=8cm]{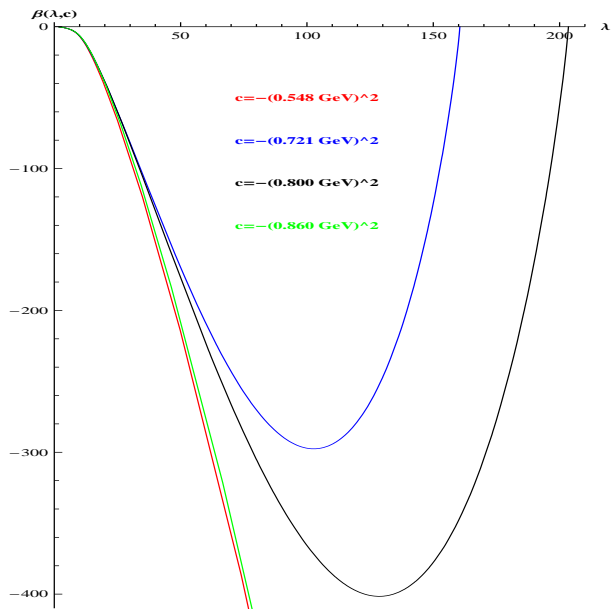}
\caption{\label{betaintermsoflambdanf6.eps} beta functions $\beta(\lambda,c)$ in terms of $\lambda$ for $\lambda_0<\lambda^{(limit)}$ and different values of $c$ in the non-extreme case when $k_1^2<16 k_2$.}
\end{center}
\end{figure}

Finally, we plot the beta functions in terms of $\lambda$ in Figure \ref{betaintermsoflambdanf6.eps}. 
Within the interval (\ref{interval}), $\beta(\lambda)$ shows the typical behaviour associated with 
an IR fixed point at finite coupling. On the other hand, 
for $c$ close to its extremal values, 
we recover the monotonous decreasing of $\beta(\lambda)$.   

%%%%%%%%%%%%%%%%%%%%%%%%%%%%%%%%%%%%%%%%%%%%%%%%%%%%%%%%%%%%%%%%%%%%%%%%

\section{Discussions and Conclusions}

The AdS/CFT correspondence provides a well-defined relation between the dimensions of Yang-Mills theory operators and the bulk masses of the corresponding supergravity fields. In the AdS/QCD models, one in general assumes that the 
same relation holds. Here we have followed this approach. 
For QCD-like gauge theories, the dimensions of operators receive anomalous contributions coming from quantum effects.
In this paper, we have developed an extended AdS/QCD approach taking into account the anomalous dimensions of the operators entering the expression of the QCD trace anomaly. We considered the scalar glueball spectroscopy within the AdS/QCD Soft-Wall framework. Following this extended model, the mass of the scalar bulk field acquires a dependence on the holographic coordinate. This modification is then reflected in the eigenvalue equation that determines the $4d$ glueball masses.

In Chapter 3, we considered some possible non-perturbative QCD beta function models discussed in the literature and calculated the corresponding glueball mass spectra, comparing with the isotropic lattice results. In particular, a beta function with an asymptotic IR cubic behaviour gives rise to the formation of a hard-wall in the $5d$ effective potential. 

In Chapter 4, we started from the linear Regge trajectory calculated using glueball lattice results and found out beta functions that lead to this spectrum.  
We find different types of beta functions classified according to their IR behaviour. 
For a class of beta functions, an IR cutoff $z_{max}$, which is not present in the Soft-Wall model, naturally emerges and gives rise to asymptotically decreasing $\beta(\lambda)$.
There is also another class of beta functions that vanish at large $\lambda$, leading to an IR fixed point at finite coupling. The IR vanishing beta functions offer a large range of possible values for the IR fixed coupling.

In quenched QCD there is only one dimensionful parameter, the QCD scale $\Lambda_{QCD}$. 
In our discussion we are also considering the quenched approximation. 
Although apparently we have two dimensionful parameters: $c$ and $ z_0$, actually, we only need one parameter as the input of the model. This could be either  $c$ or $ z_0$. From the practical point of view of the numerical calculations we fixed $z_0$ from the beginning and then obtained $c$ as an output of the model, from the best fit of the masses. So, we can redefine $c$ in terms of a dimensionless parameter $\bar c = c z_0^2 $. For the fixed value of $ z_0 = 1$ GeV$^{-1}$
we found for the three beta functions considered in section {\bf 3.2}  values for $\lambda_0$ close to each other: 18.5 , 19 and 17.9. We also found similar values for $c$: -0.35 , -0.25  and -0.30. 
For these beta functions the mass matching condition implies that $c$ is negative in our model.  

Note that in the original soft wall model \cite{Karch:2006pv,Colangelo:2007pt} the masses, shown in Eq. 
(\ref{masssoftwall}),  are proportional to the dimensionful parameter $\sqrt{c}$. Then, the mass ratios are explicitly independent of this parameter. In the model proposed here the situation is not so simple.
The inclusion of the anomalous dimension contribution in the potential, as shown in 
Eq. (\ref{potencialanomalous}), leads to a non trivial dependence of the potential on $z_0$ through the beta function.  

The QCD scale $\Lambda_{QCD}$ can be defined as the energy scale where the perturbative coupling constant diverges. So, we could estimate this scale in our model as the scale related to $z_{max}$ that appears for the beta function of section {\bf 3.2.3 } finding $\Lambda_{QCD} \approx 0.3 $ GeV. For the other beta functions of  section {\bf 3.2} this scale  could be estimated from the value of $\sqrt{\vert c \vert }$. In this case we find $\Lambda_{QCD} \sim 0.5 - 0.6 \,\,$GeV.

A better knowledge of glueball mass spectroscopy would further constrain the holographic parameters. This could help answering the question of the presence or not of an IR fixed point and its value.  

  It would be interesting to extend this procedure to other particles to investigate further the relation between the bulk mass and the beta functions. From the expression of the trace anomaly  of the QCD energy-momentum tensor in Eq. (\ref{traceanomaly}) one sees that it is possible to apply a similar method to scalar mesons, represented by the operator $\bar{q} q$. In particular, for this operator, following a procedure analogous to that of section {\bf 2.2}, one finds:
\begin{equation}
\Delta_{\bar{q}q}=3+\frac{\beta(\alpha)\gamma_m^{\prime}(\alpha)}{(1+\gamma_m(\alpha))}-\gamma_m(\alpha)
\end{equation}
\noindent with the anomalous dimension of the (light) quarks 
$\gamma_m(\mu)\equiv-\frac{d\ln m_q(\mu)}{d\ln\mu}$.  This problem will be investigated in the future.

\bigskip

\noindent {\bf Acknowledgments:} The authors are partially supported by Capes and CNPq, Brazilian agencies. One of us, F.J., is grateful to J. A. Helay\"el-Neto for his hospitality at the CBPF during the completion of this work.

 \end{document}